# Evidence of a cluster spin-glass state in the B-site disordered perovskite $SrTi_{0.5}Mn_{0.5}O_3$

Shivani Sharma[2,3*], Poonam Yadav[1], Tusita Sau[1], Premakumar Yanda[2], Peter J. Baker[3], Ivan da Silva[3], A. Sundaresan[2], N. P. Lalla[1]

[1]*UGC-DAE Consortium for Scientific Research, Indore, India-452001*
[2]*Chemistry and Physics of Materials unit, Jawaharlal Nehru Center for Advanced Scientific Research, Bengaluru, India-560064*
[3]*ISIS Facility, Rutherford Appleton Laboratory, Chilton, Didcot OX11 0QX, United Kingdom*

## Abstract

$SrTi_{0.5}Mn_{0.5}O_3$ (STMO) is a chemically disordered perovskite having random distribution of Ti and Mn over 1b site. Striking discrepancies about the structural and magnetic properties of STMO demands detailed analysis which is addressed. To explore the magnetic ground state of STMO, static and dynamic magnetic properties were studied over a broad temperature range (2-300 K). The *dc*, *ac* magnetization show a cusp like peak at $T_f \sim$ 14 K, which exhibits field and frequency dependence. The thermoremanent magnetization is characterized by using stretched exponential function and characteristic time suggests the existence of spin clusters. Also the other features observed in magnetic memory effect, muon spin resonance/rotation and neutron powder diffraction confirm the existence of cluster spin glass state in STMO, rather than the long range ordered ground state. Intriguingly, the observed spin relaxation can be attributed to the dilute magnetism due to non-magnetic doping at Mn-site and competing antiferromagnetic and ferromagnetic interactions resulting from the site disorder.

## I. Introduction

Competing magnetic interactions added to chemical disorder can hinder long range magnetic ordering and may lead to frozen spin states at low temperatures. The key point of present work is to understand such effect of chemical disorder on magnetic ground state of B-site of double perovskite $SrTi_{0.5}Mn_{0.5}O_3$ (STMO). Meanwhile, another important factor which is found to promote the spin-glass (SG) state in double perovskites is geometrical frustration arising from the tetrahedral topology[1,2,3] of the magnetic ion. Here, the term double perovskite has been specifically used for B-site ordered systems having the general formula $A_2BB'O_6$. In these systems, geometrical frustration does not always lead to the short range ordered state but can instead produce some more exotic magnetic ground states including spin-ice, spin-liquid etc.[4,5,6], in addition to long range order. Intriguingly, for B-site ordered double perovskites having magnetic cation at B'-site, chemical disorder and geometrical frustration appear to be mutually exclusive. In a B-site ordered magnetic double perovskites, the magnetic ions form the three dimensional edge sharing tetrahedral topology, which gives rise geometrical frustration in face-centered cubic symmetry[3]. On the other hand, in the case of chemically disordered arrangement, there will be a multiplicity of magnetic interaction pathways in the lattice. Therefore, such tetrahedral topology could not be expected to induce short-range spin order but indeed one can never rule out the possibility of having geometrically frustrated local regions.

For the present study, 50% non-magnetic ions ($Ti^{4+}$) doped $SrMnO_3$ (SMO) is chosen to understand the role of chemical disorder in governing the magnetic properties. Pristine SMO is reported to form either cubic (high temperature) or hexagonal (4H or 6H low temperature) phase depending on the synthesis conditions. Irrespective of the structure, SMO undergoes antiferromagnetic (AFM) ordering below room temperature with $T_N$ slightly differing for different structures[7,8,9]. The structural and magnetic properties of B-site disordered STMO have been examined through X-ray diffraction (XRD), magnetization, heat capacity, memory effect, thermoremanent magnetization (TMR), muon spin relaxation/rotation (μSR) and neutron powder diffraction (NPD) measurements. In existing literature, STMO has been assigned a double perovskite *Fm*-3*m* structure with chemical formula $Sr_2TiMnO_6$[10,11,12,13]. Due to almost similar size and isovalent state of $Mn^{4+}$ and $Ti^{4+}$ ions, it is difficult to synthesize chemically ordered STMO under normal pressure condition[14]. However, recently the ordering of isovalent cations $Fe^{3+}$ and $Al^{3+}$ was reported in $Bi_2FeAlO_6$ under high pressure[15]. Our previous work on STMO was dedicated to explore the detailed structural properties using XRD and transmission electron microscopy which confirmed that STMO forms a chemically disordered structure with random distribution of $Mn^{4+}$ and $Ti^{4+}$ ions over 1*b*-site in a cubic (*Pm*-3*m*) structure[16,17]. The presence of a single anomaly at ~14 K in magnetization was also observed, which due to lack of microscopic evidence was previously attributed to some long range ordering of $Mn^{4+}$ spin. Indeed, there are several reports on the magnetic properties of STMO including NPD[10,11,12,13,18] data but to the best of our knowledge, none of them provide sufficient microscopic information to reveal the true magnetic ground state. Noticeably, the previous NPD studies were performed on STMO having magnetic impurities and therefore, the magnetic peaks observed in NPD below 14 K were possibly misinterpreted[12]. In the present work, extensive experimental efforts to reveal the microscopic evidence of magnetic ground state of STMO are presented. Our detailed *ac* and *dc* magnetization studies including the results of microscopic probes like, μSR, memory effect, thermoremanent magnetization and NPD confirm the cluster SG state of STMO, in contrast to A-type AFM state[12].

## II Experimental details

The polycrystalline sample of $SrTi_{0.5}Mn_{0.5}O_3$ (STMO) was prepared through a solid state reaction route using $SrCO_3$, $TiO_2$ and $MnO_2$, following the procedure described in Ref. 16. Powder XRD data in the $2\theta$ range of 10-100° were recorded using Cu-$K_\alpha$ X-rays to confirm the phase formation. Rietveld refinement of laboratory powder XRD pattern confirms the single phase *Pm*-3*m* space group having random distribution of $Ti^{4+}$ and $Mn^{4+}$ over 1*b*-site using the previously mentioned structural parameters[17]. The temperature, field and time dependent *dc*-magnetization measurements were carried out using SQUID MPMS3 under various protocols like zero field cooled (ZFC) and field cooled (FC) conditions, with cooling and heating rates of 3 K /min. The temperature dependent NPD and zero field (ZF) μSR experiments were carried out using GEM and MuSR instruments at the ISIS Neutron and Muon source, UK. Heat capacity measurement was carried out in Physical Property Measurement System (PPMS) by Quantum Design using relaxation method. Rietveld analysis of XRD and NPD data has been performed to extract the structural information using JANA2006[19].

## III Results and Discussion

### A. Physical properties:

Temperature dependent *dc*-magnetic susceptibility $\chi_{dc}(T)$ of STMO measured with 0.01 and 1 Tesla magnetic fields under ZFC and FC conditions is presented in figure 1. For 0.01 Tesla, ZFC exhibits a sharp peak below 14 K with bifurcation between ZFC and FC curves. Intriguingly, the bifurcation exist even above 14 K up to ~ 40 K which may be due to the present of very small impurity of $Mn_3O_4$ present in the sample, unlike our previous report on same composition[16]. It is quite possible that a very small magnetic impurity of $Mn_3O_4$ may remain undetected in diffraction experiments but may show some features in more sensitive magnetization measurements. Meanwhile, the $\chi_{dc}(T)$ decreases at 1 Tesla and the peak in ZFC shifts towards lower temperature along with the decrease in bifurcation. Also, the peak becomes broader in 1 Tesla field, as compared to 0.01 Tesla. The peak temperature in $\chi_{dc}(T)$ appears due to freezing of the moments and therefore is usually denoted by $T_f$, the freezing temperature. Shift of the freezing-transition ($T_f$), broadening of peak and decrease in $\chi_{dc}(T)$ with increasing field, all together infer the glassy behavior of STMO below 14 K[20]. The value of Weiss constant ($\theta_{CW}$) as obtained from *CW*-fit, comes out to be negative, around -382 K, see inset of figure 1, which implies that below 160 K, the dominating interaction is AFM. Figure 2 exhibits the isothermal magnetization (*M*-vs-*H*) curves measured at 3, 10, 16, 100 and 300 K. For 3 K, a small and clear characteristic "S" shape hysteresis is observed with coercive field of ~2000 Oe and very small magnetization at 7 T. With increasing temperatures, the "S" shape curves transform into the linear curve, as expected from paramagnetic (PM) state. The similar kind of curves for STMO were reported earlier and have been attributed to AFM behavior with weak FM component[10,11,12,13,16,18]. Noticeably, in most of these studies except Sharma et al.[16] and Qasim et al.[18], an additional transition was reported at 45 K. In our previous work, presence of 45 K transition was ruled out and it was attributed to a $Mn_3O_4$ impurity[16]. Therefore, the observed *M*-vs-*H* behavior is attributed to the competing ferromagnetic and AFM interactions present in the system leading to SG state. Similar M-H behavior is reported for many other SG systems[20,21,22,23].

To understand the dynamics of the SG state, temperature dependent *ac*-magnetic susceptibility $\chi'_{ac}(T)$ was also measured at different frequencies with fixed excitation amplitude of 3 Oe and presented in figure 3(a) over the 2-35 K temperature range. A small shift in peak with frequency following by a dispersion at low T around ~13.2 K is observed, as usually expected for SGs[20,23,24]. An enlarged view is presented in the inset to show the presence of frequency dispersion, however, the frequency dispersion of the maxima is very small and difficult to visualize. But in lower temperature regime (2-8 K), the dispersion enhances, which can be seen in the encircled region. Thus the ac-susceptibility data confirms the SG behavior of STMO below 14 K.

Figure 3(b) represents the heat capacity results, in which $C_p/T$ is plotted against temperature ($T$) from 5 to 35 K. A broad hump spanning from 4 to 15 K can be observed. The inset shows the $C_p/T$ variation as a function of $T^2$ close to 14 K, exhibiting clear change in the entropy. Similar features in the heat capacity were also observed for other SG system[23]. Our previous study on STMO discerns some more detail analysis of heat capacity data to estimate the magnetic entropy change over a broad temperature regime which infers that the short range correlations in STMO develops at temperatures as high as 160 K and most of the entropy is already frozen out in short range ordering above $T_f$ and therefore, only a small anomaly is observed near 14 K[16]. Therefore, the heat capacity and magnetization results support the SG state in STMO.

**B. Memory and relaxation measurements:**

To understand the non-ergodicity of spin dynamics, waiting time memory experiments were carried out in ZFC condition at different temperatures, namely at 4, 8, 12 and 16 K. For these measurements, the sample was first cooled from 300 K (PM region) down to the desired temperature and then the cooling was interrupted for 4 hrs. After waiting for 4 hrs. at that temperature, the cooling was resumed and the sample was cooled down to 2 K. Then at 2 K, 100 Oe field was applied and the magnetization ($M_{ZFC}^{mem}$) was measured as a function of temperature during heating. The same protocol was followed for all temperatures with same waiting time of 4 hrs. The measurement at 16 K was performed to confirm the presence of short range correlations above SG ordering. The results of memory experiments are depicted in figure 4 in which the reference ZFC data ($M_{ZFC}$) without any waiting time is plotted with ($M_{ZFC}^{mem}$). As expected, a clear dip can be observed for 4, 8 and 12 K. For 16 K $M_{ZFC}^{mem}$, no dip is observed as expected but after comparing this with $M_{ZFC}$, it is found that the overall magnetization remains small up to 80 K and it overlaps with $M_{ZFC}$ only at higher temperatures by restoring the previous magnetization. This also signifies the presence of short-range correlations above 14 K with the evolution of SG state below $T_f$.

In SGs, some of the magnetized regions respond much more slowly than the others and therefore, thermoremanent magnetization (TMR) decay measurements have been performed under ZFC and FC conditions to characterize the relaxation behavior and rate in the vicinity of $T_f$. For ZFC curves, the sample was cooled from 300 K (PM region) down to desired temperature and then 100 Oe field was applied to record the magnetization $M(t)$ as a function of time (t). For FC $M(t)$ curves, the sample was cooled in the presence of field H = 100 Oe down to desired temperature and then the field was removed and $M(t)$ was recoded as function of time. Following these procedures, data has been collected at 2, 8, 12 and 20 K during ZFC and FC protocols and for each dataset, the sample was cooled from PM state and data was recorded for 3 hrs. The combined results of ZFC and FC measurements are presented in figure 5 (a,b). The time dependence of TMR i.e. *M(t)* could be fitted with stretched exponential function

$$M(t) = M_0 - M_g\left[-\left(\frac{t}{\tau}\right)^{\beta}\right] \quad (1)$$

where $M_0$, $M_g$ are the intrinsic and glassy component of magnetization whereas $\tau$ and $\beta$ are the characteristic time constant and critical exponent respectively[25]. For many SGs, the reported value of $\beta$ lies between 0 and 1[20,23,26,27]. For STMO, both ZFC and FC data has been fitted with the above equation and the value of $\tau$ and $\beta$ for different temperatures are mentioned in respective figures. The values of $\tau$ for 12 K ZFC and FC data are 748 and 419 secs which correspond to the slow relaxation of spin, behaving like the cluster SG[20,23,28]. Further, with lowering the temperature, the relaxation dynamics get slower as a result of enhanced spin freezing and therefore the value of $\tau$ increases with decreasing temperature. This is in full agreement with the memory-effect experiment. This is also evident from the *ac*-susceptibility data where the frequency dispersion increases below 8 K due to frozen spin dynamics. The relaxation mechanism at 2K is very slow, and the value of $\tau$ approaches 30 ks under ZFC and 62 ks under FC condition. Our detailed relaxation studies thus confirm the cluster SG state in STMO below ~ 14 K.

**C. μSR and Neutron diffraction studies:**

Zero Field (ZF) μSR provides a unique opportunity of analyzing the SG state because in this measurement, no external field is involved which can perturb the system, unlike the magnetic susceptibility. Furthermore, μSR probes shorter timescales and a much wider relaxation time domain compared to $\chi'_{ac}(T)$. Therefore, ZF μSR measurements were carried out to investigate the cluster SG state between 40-2 K using 100% spin-polarized $\mu^+$ beam. Figure 6 (a) exhibits the ZF $\mu$SR asymmetry spectra $A(t)$ of STMO at some selected temperatures to avoid the haziness in data presentation. For each temperature curve, see the supplementary information**Error! Bookmark not defined.**. An exponential relaxation added to a background ($A_2$) was used to fit the $A(t)$ spectra.

$$A(t) = A_1\, exp(-\lambda t) + A_2 \tag{2}$$

where $A_1 + A_2$ is the asymmetry fitted from the earliest measured time in which $A_1$ is the component showing spin relaxation and $\lambda$ is the muon spin relaxation rate arising mainly due to development of local field at the muon site during freezing of the moments and also from the spin fluctuations present at all temperatures[29,30,31]. However a component of $\lambda$ may also be due to spin-fluctuations present at all temperatures. The variation of $A_0$ and $\lambda$ with temperature is presented in figure 6(b). The $\lambda(T)$ shows a clear peak around 11.5 K and $A_0$ also changes rapidly below 17 K and reaches the lower values below 11.5 K. Since the muon spin relaxation is likely to be dominated by the short-ranged dipolar fields from nearby magnetic moments, these observations reflect the microscopic nature of the spin freezing taking place at low temperatures in STMO.

To strengthen our observation of SG state and to eliminate the possibility of any long range magnetic ordering, temperature dependent NPD measurements were performed at the GEM neutron diffraction instrument of ISIS facility, UK. Figure 7(a) exhibits the stack of bank-3 data sets of time-of-flight (TOF) NPD data of STMO taken across the freezing temperatures, namely at 30, 11 and 5 K. As can be clearly noticed, we could not find evidence of any type of magnetic ordering, long or short whatsoever, below the freezing temperature, at 11K or 5K. The 5 K NPD data was Rietveld refined using *Pm*-3*m* space group. The refined NPD pattern is presented in figure 7(b) and the temperature variation of lattice parameter is plotted in the inset. The structure of STMO remains cubic down to 5 K. Thus the cusp like peaks in *dc* and *ac* $\chi(T)$, the corresponding peak in muon-spin relaxation parameter ($\lambda$) and the absence of any signature of long range AFM ordering clearly prove the occurrence of spin freezing in STMO.

## IV. Conclusions

The *dc*-magnetization reveals a cusp like peak at $T_f \sim$ 14 K and its field dependence indicates a glassy behavior. Results of TMR and memory-effect using waiting time protocol, affirm spin-relaxation in STMO below $T_f$, supported by the heat capacity. The frequency dispersion of the cusp-like anomaly at 14 K in *ac*-susceptibility and the slow relaxation, as probed through TMR measurement, manifest a cluster spin-glass state in STMO. Also, the increase in relaxation time with decreasing temperature is interpreted in terms of slower relaxation dynamics at low temperature (below 8 K). The occurrence of sharp cusp like peak in muon spin-relaxation rate $\lambda(T)$ near $T_f$ , confirm the spin-freezing in STMO. Rietveld refinement of NPD patterns confirms that the nuclear structure remains cubic down to 5 K and the absence of any new peak across the spin-freezing transition, confirms absence of any long range magnetic ordering i.e. presence of a spin-glass state in STMO. The dilute magnetism due to non-magnetic doping at $Mn^{4+}$ site and competing AFM and ferromagnetic interactions are responsible for the observed spin-glass state in STMO. The non-magnetic doping at $Mn^{4+}$ site strongly affect the superexchange interactions between magnetic cations which results in the glassy state, unlike the pristine SMO which is a long range order AFM. Based on the above described magnetic, μSR and NPD studies on 50% B-site disordered Ti substituted $SrMnO_3$ we conclude that STMO undergoes a cluster spin-glass transition at ~ 14 K.

## V. Acknowledgement

Dr. Shivani Sharma gratefully acknowledges Department of Science and Technology, India for providing postdoctoral fellowship through Indo-UK nanomission project. Authors are thankful to ISIS neutron and muon source, UK for providing experimental support to perform μSR and neutron diffraction measurements. Authors are also thankful to Dr. Alok Banerjee and Dr. R. Rawat for providing experimental support for *ac*-susceptibility and specific heat measurements.

**Figure caption:**

**Figure 1:** Temperature dependent ZFC and FC $\chi_{dc}(T)$ measured with 0.01 and 1 Tesla. The inset shows the $CW$-fit of inverse susceptibility data.

**Figure 2:** Isothermal magnetization curves at various temperatures. The inset presents an enlarged view of the 2 K curve.

**Figure 3:** (a) $\chi'_{ac}(T)$-vs-$T$ behavior of STMO at different frequencies. Inset shows the enlarge view (i) around $T_f$ and (ii) below 8 K. (b) Temperature dependent heat capacity behavior in the absence of magnetic field. The $C_p/T$ vs $T^2$ behavior around $T_f$ is plotted in the inset.

**Figure 4:** Memory effect as a function of temperature in ZFC conditions with 100 Oe field at 4, 8, 12 and 16 K ($M_{ZFC}^{mem}$) with reference data ($M_{ZFC}$). For each $M_{ZFC}^{mem}$ data set, the waiting time was 4 hrs. The insets show (a) the enlarged view between 2 and 18 K , showing clear dips in $M_{ZFC}^{mem}$ curve for 4, 8 and 12 K, (b) and (c) exhibit the difference and overlapping between $M_{ZFC}$ (black) and $M_{ZFC}^{mem}$ (red, 4 K) curves in intermediate and high temperature regime.

**Figure 5:** Relaxation of (a) ZFC and (b) FC magnetization as a function of time at different temperatures. The solid red line indicates the stretched exponential fit using equation (1).

**Figure 6:** (a) Temperature dependent ZF $\mu$SR asymmetry spectra for STMO at selected temperatures between 40 and 2 K. The red line shows the least square fit of the data using equation (2). **(b)** Temperature variation of the asymmetry $A_0$ and the relaxation rate $\lambda$ as obtained from the *A(t)-vs-t* data, using equation.(2).

**Figure 7:** (a) Time-of-flight NPD patterns recorded at 30, 11 and 5 K. The inset shows the zoomed view at higher d-spacings. (b) Rietveld refinement of 5 K NPD pattern using *Pm*-3*m* space group. The inset shows the temperature variation of the lattice parameter. The goodness of fit parameters are: $\chi^2$ =2.41, $R_p$ =3.05 , $R_{wp}$ = 4.80.

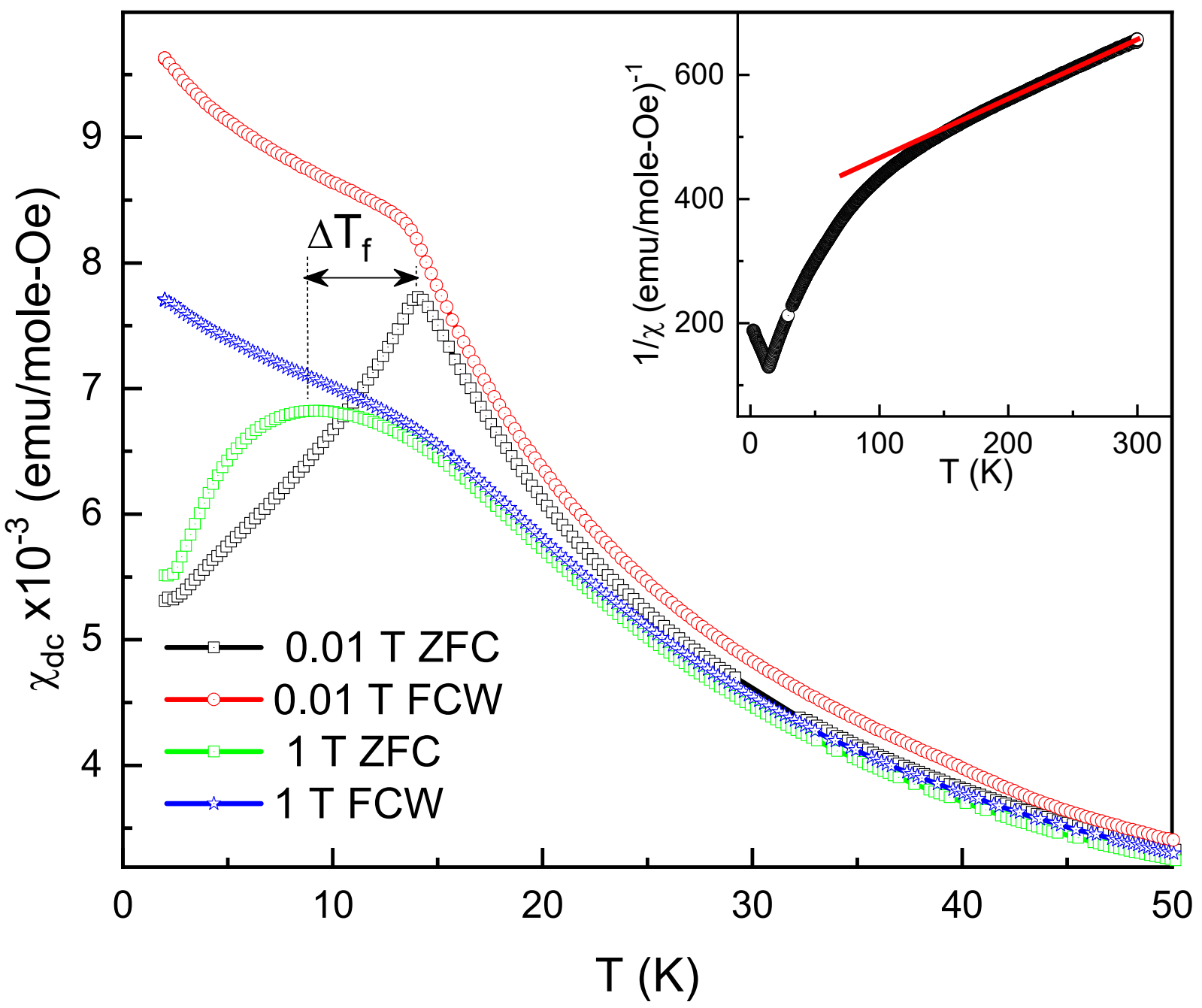

**Figure 1**

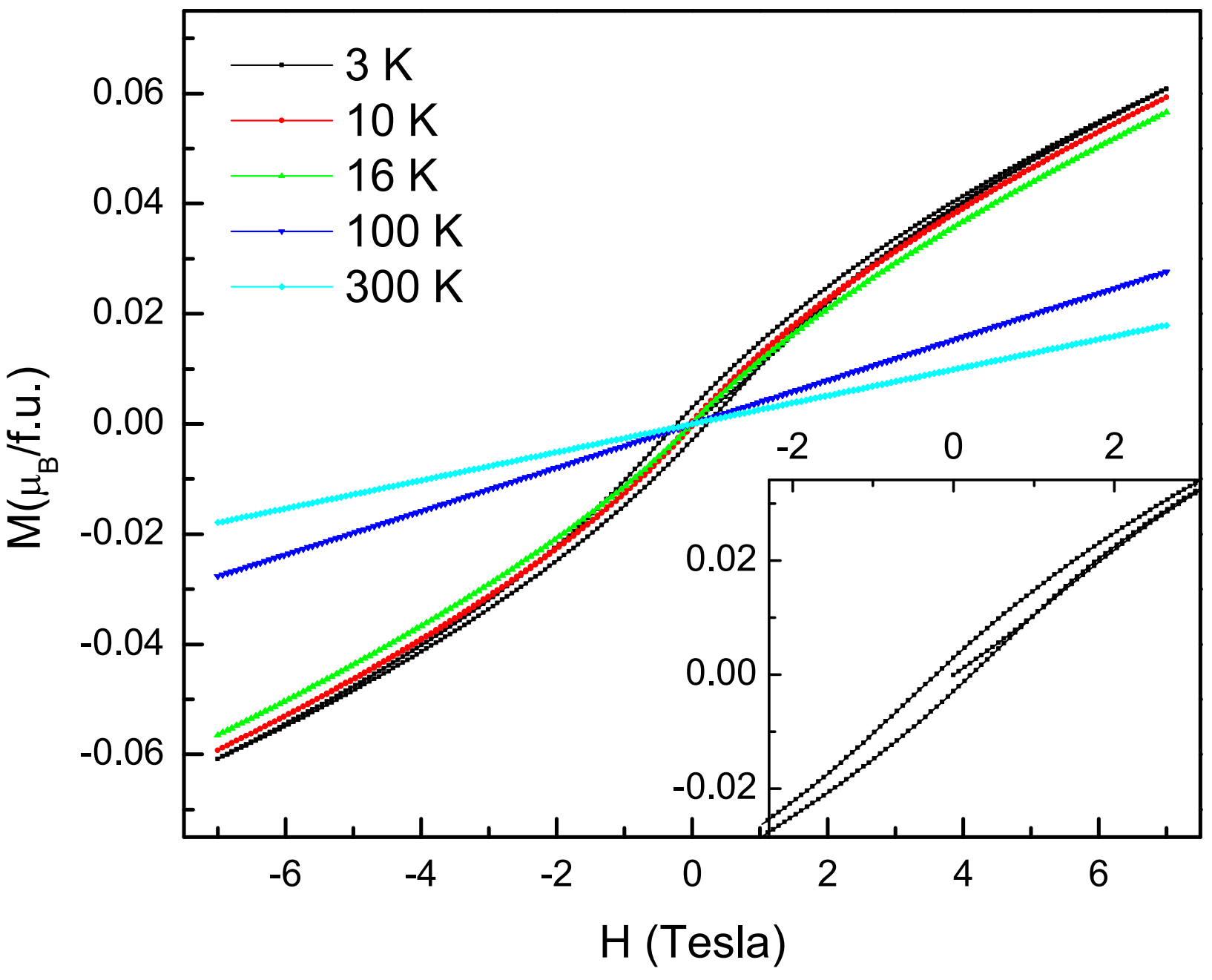

**Figure 2**

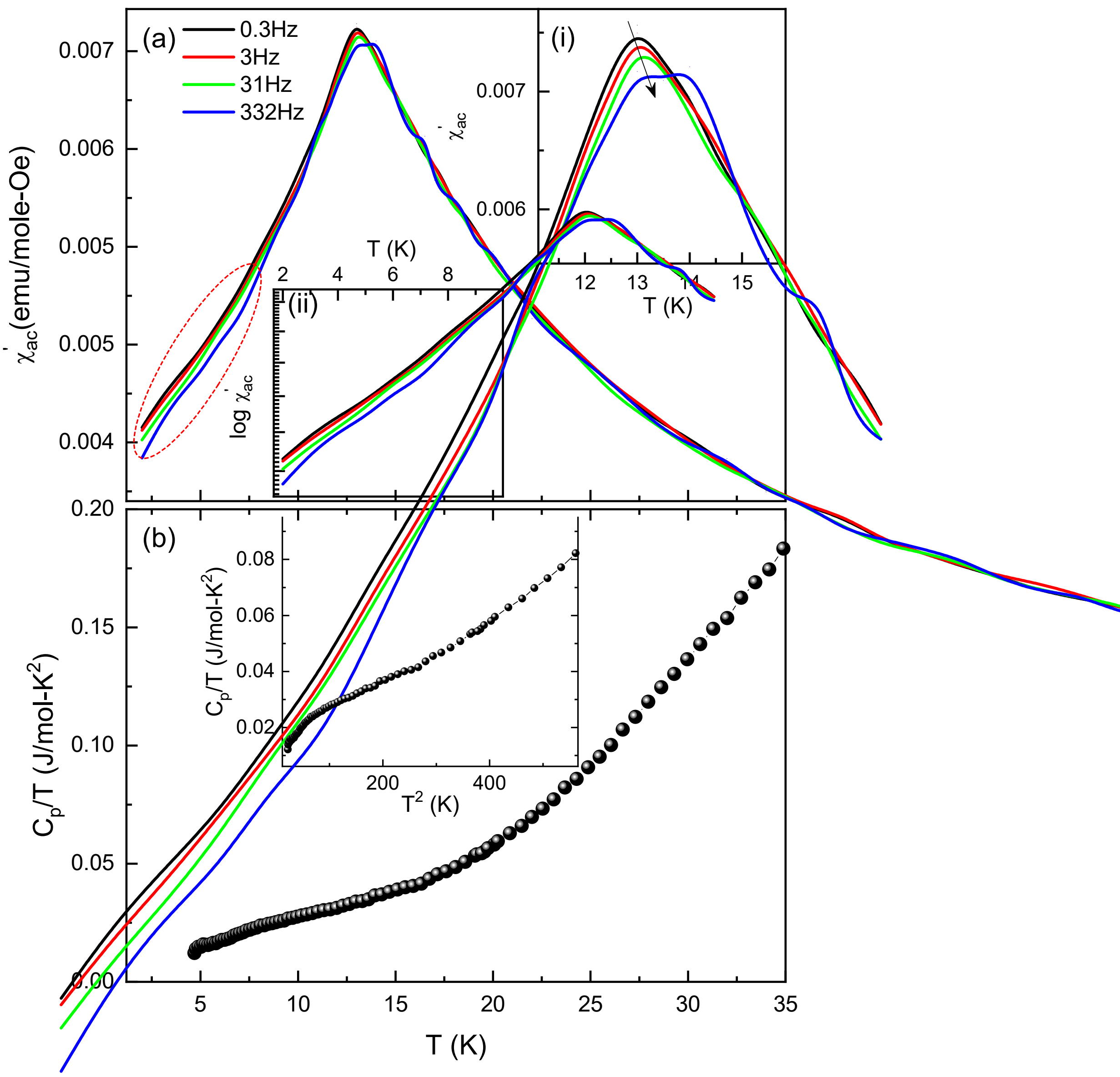

**Figure 3**

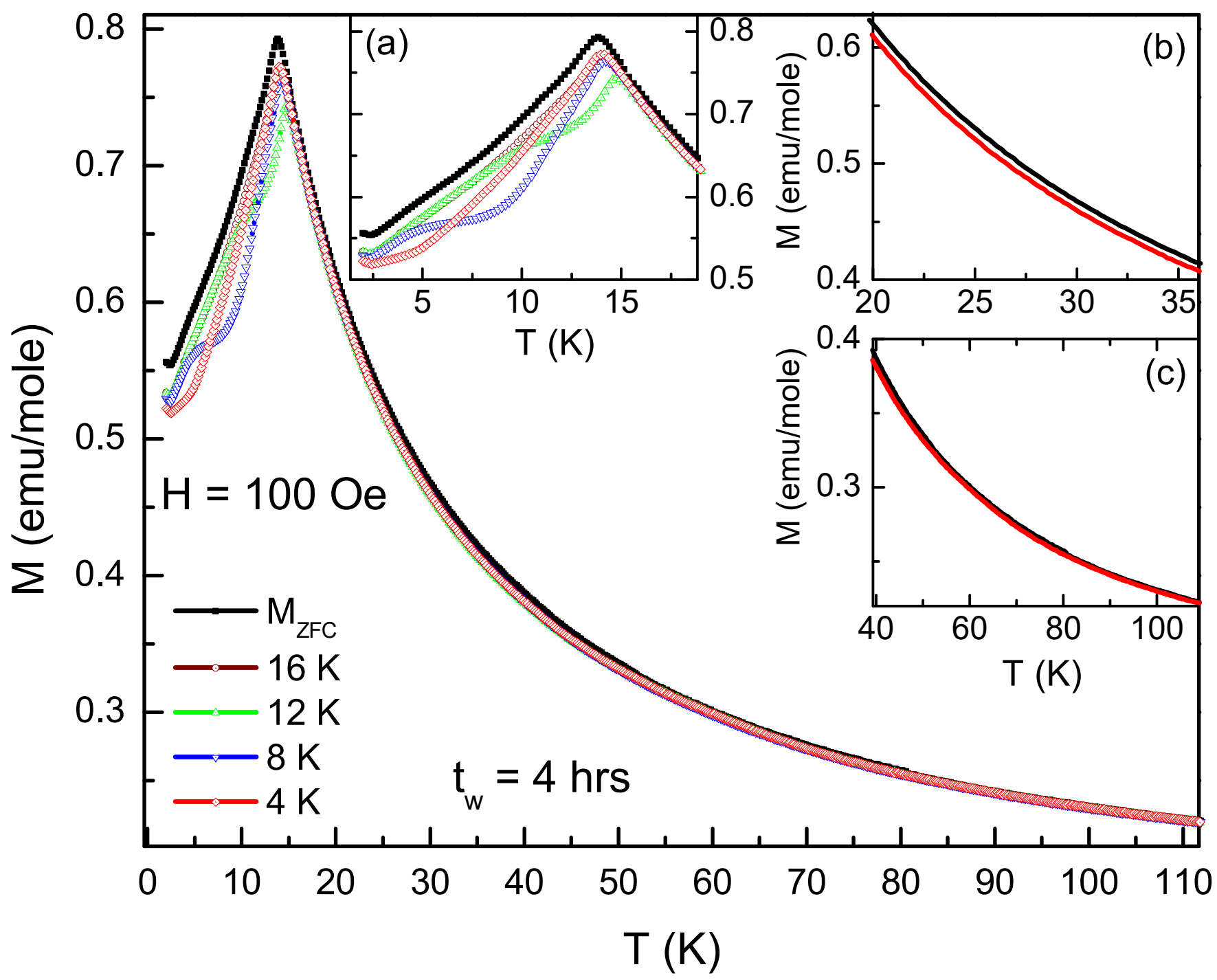

**Figure 4**

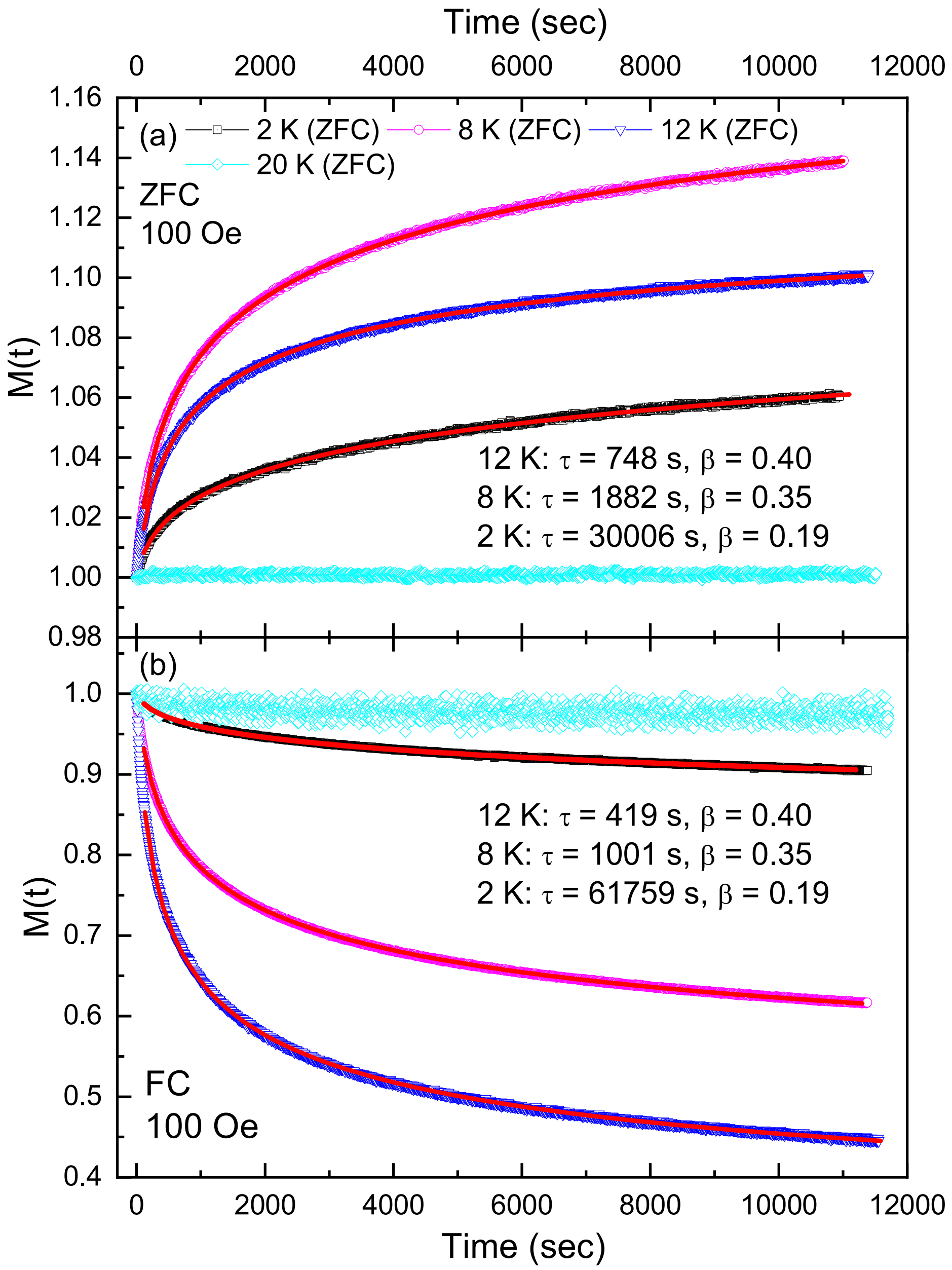
Time (sec)
0
2000
4000
6000
8000
10000
12000
1.16
1.14
1.12
1.10
1.08
1.06
1.04
1.02
1.00
0.98
M(t)
(a)
2 K (ZFC)
8 K (ZFC)
12 K (ZFC)
20 K (ZFC)
ZFC
100 Oe
12 K: τ = 748 s, β = 0.40
8 K: τ = 1882 s, β = 0.35
2 K: τ = 30006 s, β = 0.19
(b)
1.0
0.9
0.8
0.7
0.6
0.5
0.4
M(t)
12 K: τ = 419 s, β = 0.40
8 K: τ = 1001 s, β = 0.35
2 K: τ = 61759 s, β = 0.19
FC
100 Oe
0
2000
4000
6000
8000
10000
12000
Time (sec)

**Figure 5**

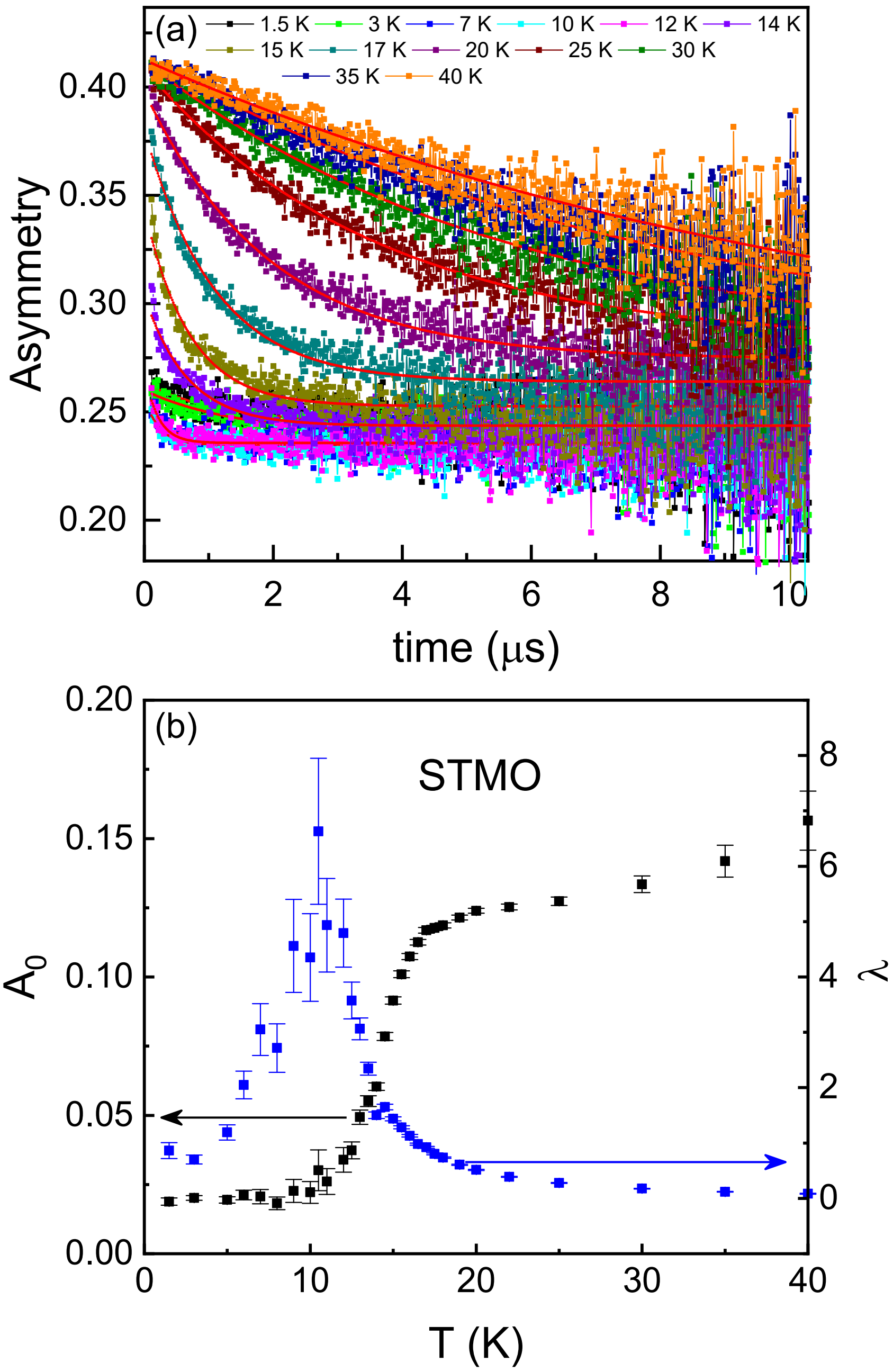

**Figure 6**

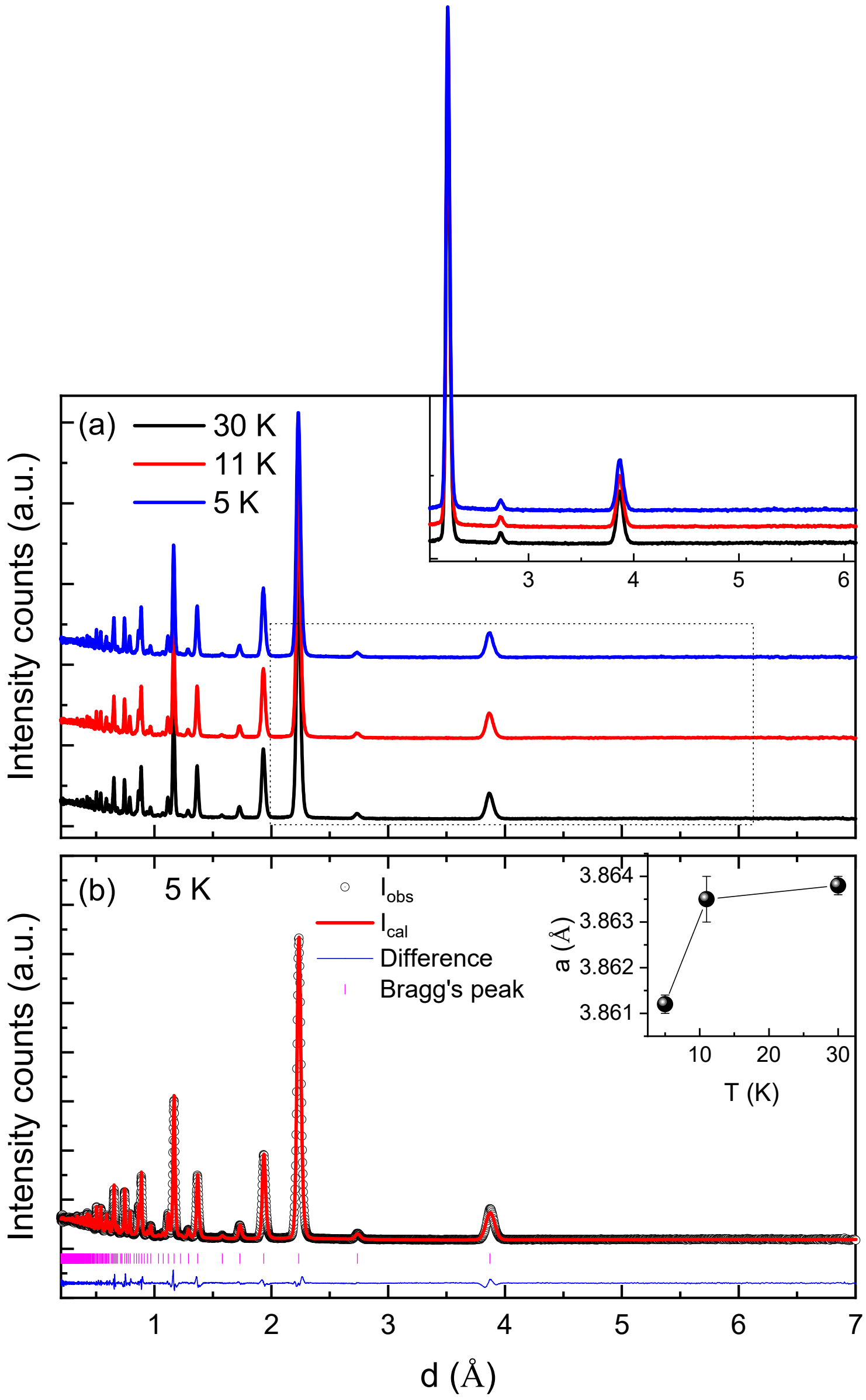

**Figure 7**